\documentclass[12pt,american,english]{article}
\usepackage[T1]{fontenc}
\usepackage[latin9]{inputenc}
\usepackage{geometry}
\usepackage{color}
\usepackage{float}
\usepackage{units}
\usepackage{textcomp}
\usepackage{amsmath}
\usepackage{amssymb}
\usepackage{graphicx}
\usepackage{esint}

\makeatletter
\@ifundefined{date}{}{\date{}}
 \renewcommand{\section}{\@startsection{section}{1}{\z@}%
                                   {-3.5ex \@plus -1ex \@minus -.2ex}%
                                   {2.3ex \@plus.2ex}%
                                   {\normalfont\normalsize\bfseries}}
 \renewcommand{\subsection}{\@startsection{subsection}{1}{\z@}%
                                   {-3.5ex \@plus -1ex \@minus -.2ex}%
                                   {2.3ex \@plus.2ex}%
                                   {\normalfont\normalsize\bfseries}}
 \renewcommand{\subsubsection}{\@startsection{subsubsection}{1}{\z@}%
                                   {-3.5ex \@plus -1ex \@minus -.2ex}%
                                   {2.3ex \@plus.2ex}%
                                   {\normalfont\normalsize\bfseries}}

\newcommand{\etal}{{\it et al}}

\usepackage{babel}

\usepackage{babel}

\makeatother

\usepackage{babel}
\begin{document}

\title{\textbf{\Large Resolution enhancement in neural networks with dynamical
synapses}}

\author{{\normalsize C. C. Alan Fung\textsuperscript{1}, He Wang\textsuperscript{1},
Kin Lam\textsuperscript{1,\dag{}}, K. Y. Michael Wong\textsuperscript{1,{*}},
Si Wu\textsuperscript{2,{*}}}}

\maketitle
\noindent \textsuperscript{1}Department of Physics, The Hong Kong
University of Science and Technology, Clear Water Bay, Hong Kong,
China

\noindent \textsuperscript{2}State Key Laboratory of Cognitive Neuroscience
and Learning, Beijing Normal University, Beijing 100875, China

\noindent \textbf{\ }\\
 \textbf{Correspondence:}\\
 Prof. K. Y. Michael Wong\\
 Department of Physics\\
 The Hong Kong University of Science and Technology\\
 Clear Water Bay\\
 Hong Kong, China\\
 phkywong@ust.hk\\
 \\
 Prof. Si Wu\\
 State Key Lab of Cognitive Neuroscience \& Learning\\
 Beijing Normal University \\
 Beijing 100875\\
 China\\
 wusi@bnu.edu.cn\\
 \\
 \textsuperscript{\dag{}}Current Affiliation: Department of Physics,
University of Illinois at Urbana-Champaign, Urbana, IL 61801-3080,
U. S. A.\\
 \\
 \textbf{Keywords}: Continuous \foreignlanguage{american}{attractor}
neural network, neural field model, short-term synaptic depression,
short-term synaptic plasticity, transparent motion\\
 Number of words: 5832\\
 Number of figures: 15 
\begin{abstract}
{\normalsize Conventionally, information is represented by spike rates
in the neural system. Here, we consider the ability of temporally
modulated activities in neuronal networks to carry information extra
to spike rates. These temporal modulations, commonly known as population
spikes, are due to the presence of synaptic depression in a neuronal
network model. We discuss its relevance to an experiment on transparent
motions in macaque monkeys by Treue \etal. in 2000. They found that
if the moving directions of objects are too close, the firing rate
profile will be very similar to that with one direction. As the difference
in the moving directions of objects is large enough, the neuronal
system would respond in such a way that the network enhances the resolution
in the moving directions of the objects. In this paper, we propose
that this behavior can be reproduced by neural networks with dynamical
synapses when there are multiple external inputs. We will demonstrate
how resolution enhancement can be achieved, and discuss the conditions
under which temporally modulated activities are able to enhance information
processing performances in general.}{\normalsize \par}
\end{abstract}

\section{{\normalsize Introduction}}

An important issue in computational neuroscience is how information
is represented in the neural system. It was widely accepted that spike
rates of neurons carry information. This notion was further illustrated
in \textit{population codes}, in which the a group of neurons encode
information and even represent uncertainties therein through their
collective activities \cite{Zemel1999,Pouget2000}. Consequently,
population coding has been successfully applied to describe the encoding
of spatial and directional information, such as orientation \cite{Ben-Yishai1995},
head direction \cite{Zhang1996}, and spatial location \cite{Samsonovich1997}.
They are also used to explain information processing in the recently
discovered grid cells \cite{Fuhs2006}.

An interesting question arises, namely, whether information can be
encoded in other aspects of population coding besides spike rates.
For example, can extra information be carried by the coding if the
spikes are modulated in time, so that different spike trains modulated
differently may convey different messages even though their spike
rates appear to be the same. Given this possibility, the information
content of population coding can be much richer than its superficial
appearance as spike rates.

In this paper, we will explore the ability of population spikes to
carry information extra to spike rates. Population spikes are temporal
modulations of the population neuronal activity, and are also known
as ensemble synchronizations, representing extensively coordinated
rises and falls in the discharge of many neurons \cite{Loebel2002,Holcman2006}.
The population spikes are due to the presence of short-term depression
(STD) of the synapses, referring to the reduction of synaptic efficacy
of a neuron after firing due to the depletion of neurotransmitters
\cite{Stevens1995,Markram1996,Dayan2001}. This adds to a recently
expanding list of the roles played by STD in neural information processing.
For example, STD was recently suggested to be useful in expanding
the dynamic range of the system \cite{Tsodyks1997,Abbott1997}, estimating
the information of the pre-synaptic membrane potential \cite{Pfister2010},
and stabilizing the self-organized critical behavior for optimal computational
capabilities \cite{Levina2007}. STD was also found to be useful in
enhancing the mobility of the network state in tracking moving stimuli
\cite{Fung2012a}, and hence was recently proposed to be a foundation
of a potential anticipation mechanism \cite{Fung2012b}.

Previously, population spikes were found to be global synchronizations
of neuronal activities. However, in order for them to encode spatial
information, the population spikes that will be considered in this
paper are localized ones. We will use the case of transparent motion
as an example. This example illustrates the possibility that the modulation
by population spikes enables the neural system to refine the resolution
of direction for multiple stimuli. The prediction by the proposed
mechanism has an excellent agreement with experimental results \cite{Treue2000}.

Transparent motion is one of the most well-known experiments in the
psychophysical community. In the experiment, the stimulus usually
contains moving dots with different directions. So, there are multiple
moving directions transparently superimposed on one another. In the
nervous system, the middle temporal (MT) area was found to be responsible
for detecting moving directions of objects \cite{Maunsell1983}. Here,
it was recently found that the neurons are heterogeneous, with some
neurons responding to the pattern of moving stimuli, while others
responding to the components of composite moving patterns \cite{Rust2006}.
In 2000, Treue \etal. found that if the directions of two groups
of moving dots differ by an angle larger than the tuning width of
the neurons, the observed neuronal response profile begins to split
\cite{Treue2000}. However, subjects can still distinguish the two
directions if their difference is as small as about 10\textdegree{}
\cite{Mather1980}, while the average direction tuning width of neurons
is about 96\textdegree{}.

To resolve this paradox, Treue \etal. proposed that when the resultant
neuronal response is too board for a single direction, the perception
can identify the two directions by considering the resultant neuronal
response as a superposition of two individual neuronal responses of
each direction. However, when the two directions differ by an angle
less than the tuning width, it becomes difficult to resolve the peaks
of the two superposed responses, if the curvature of the average neural
activity profile is not taken into account This difficulty was also
observed in simulations with distributional population codes \cite{Zemel1999}.
The mechanism of enhanced resolution remained unknown, and coding
by firing rates may not reveal the complete picture.

In a recently proposed model on motion transparency, the enhanced
resolution was achieved \cite{Raudies2011}. Two mechanisms held the
key to this advance. First, as in standard neural field models, there
is a local center-surround competition in the space of motion directions.
Although this is not sufficient to explain the enhanced resolution,
there is the second mechanism, namely, the modulatory feedback signals
from higher stages of processing in the area medial superior temporal
(MST) area. Motion attraction (that is, under-estimation of the directional
difference) at small angular difference, and motion repulsion (that
is, over-estimation) at larger angles were successfully explained.
Perception repulsion can also be found in a Bayesian inference explanation
on identification of audiovisual stimulus \cite{Sato2007}.

Here, we propose a novel mechanism for resolution enhancement based
on the temporal modulation inherent in population coding. To focus
on the generic issue of whether information carried in the temporal
modulation of population coding can be usefully applied in a processing
task, we consider a simplified model of transparent motion. We assume
that inputs from different locations of the receptive field have been
integrated, the directional information has been filtered, and the
processing of input information can proceed without the assistance
of feedback modulations. Thus our working model reduces to a single
network. The working principle is a continuous attractor neural network
(CANN) with dynamical synapses. Continuous attractor neural networks,
also known as neural field models, are models used for describing
phenomena and features observed in some brain regions where localized
attractor neuronal responses are used to represent continuous information.
Due to short-range excitatory interactions and long-range/global inhibitory
interactions, bump-shaped neuronal response profiles are attractors
of CANNs. Since the response profiles are easy to shift their positions
in the space of continuous information, they are useful in tracking
moving stimuli \cite{Ben-Yishai1995,Amari1977,Wu2008,Fung2010} and
their drifting behaviors have been studied \cite{Itskov2011}. In
contrast to these studies of tracking, we will focus on stationary
stimuli and their time-dependent neuronal responses.

Dynamical synapses are found to enrich the dynamical behaviors of
CANNs \cite{York2009,Fung2012a}. Short-term synaptic depression (STD)
can degrade the synaptic efficacies between neurons, depending temporally
on the activity history of the presynaptic neuron \cite{Tsodyks1998}.
In the presence of an external stimulus, the bumps can remain temporally
stable if STD were absent. However, with STD, the population activity
may drop after it reaches a maximum, since neurotransmitters have
been consumed. After the drop, neurotransmitters are recovered and
the neuronal population is ready to respond to the external stimulus
again. This results in periodic bursts of local neuronal responses,
referred to as population spikes. As we shall see, the temporal modulation
induced by STD, together with input fluctuations, enable the system
to reduce the angle of resolution in transparent motion down to one-fourth
to one-third of the tuning width of the neuron.

In the rest of this paper, we will begin with an introduction of the
CANN model and its basic properties. After that, we will discuss simulation
results showing that our model is able to represent acute difference
in transparent stimuli. At the end, there is a discussion section
concluding our proposed mechanism.

\section{{\normalsize Model and method}}

In the continuous attractor neural network model, we specify the dynamics
and the state of the system by the neuronal current. For neurons with
preferred stimulus $x$ in the range $-L/2\leq x\leq L/2$, its neuronal
current is denoted by $u\left(x,t\right)$. The dynamics of $u\left(x,t\right)$
is given by \cite{Fung2012a} 
\begin{eqnarray}
\tau_{s}\frac{du}{dt}\left(x,t\right) & = & -u\left(x,t\right)+I^{\text{ext}}\left(x,t\right)+\rho\int dx'J\left(x-x'\right)p\left(x',t\right)r\left(x',t\right).\label{eq:dyn_u}
\end{eqnarray}
$\tau_{s}$ is the timescale of $u\left(x,t\right)$. It is usually
of the order of the magnitude of 1 ms. $\rho$ is the density of neurons
over the space spanned by $\left\{ x\right\} $. $J\left(x-x'\right)$
is a translational invariant excitatory coupling given by 
\begin{equation}
J\left(x-x'\right)=\frac{J_{0}}{\sqrt{2\pi}a}\exp\left(-\frac{\left|x-x'\right|^{2}}{2a^{2}}\right),\label{eq:Jxx}
\end{equation}
where $a$ is the range of excitatory connection and $J_{0}$ is the
average strength of the coupling. $r\left(x,t\right)$ is the neural
activity related to $u\left(x,t\right)$ by 
\begin{equation}
r\left(x,t\right)=\Theta\left[u\left(x,t\right)\right]\frac{u\left(x,t\right)^{2}}{B\left(t\right)}.\label{eq:rxt}
\end{equation}
Here, $\Theta$ is a step function centered at 0. The denominator,
$B(t)\equiv1+k\rho\int dx'u\left(x',t\right)^{2}$, in this formula
is the global inhibition, controlled by the inhibition parameter $k$.
This type of global inhibition can be achieved by shunting inhibition
\cite{Heeger1992,Hao2009}. $I^{\text{ext}}\left(x,t\right)$ is the
external input to the system, which will be defined in the latter
part of this section.

In the integral of Eq. (\ref{eq:dyn_u}), $p\left(x,t\right)$ is
the available fraction of neurotransmitters of the presynaptic neurons.
Neurotransmitters are consumed when a neuron sends chemical signals
to its postsynaptic neurons. However, the recovery time of the neurotransmitters
is considerably longer than $\tau_{s}$. This process can be modeled
by \cite{Tsodyks1998,Fung2012a} 
\begin{equation}
\tau_{d}\frac{dp}{dt}\left(x,t\right)=-p\left(x,t\right)+1-\tau_{d}\beta p\left(x,t\right)r\left(x,t\right).\label{eq:dyn_p}
\end{equation}
$\tau_{d}$ is the timescale of recovery process of neurotransmitters.
The recovery process usually takes 25-100 ms. Here, we choose $\tau_{d}=50\tau_{s}$.
These two differential equations, Eqs. (\ref{eq:dyn_u}) and (\ref{eq:dyn_p}),
are found to be consistent with the model proposed by Tsodyks \etal.
in 1998 {[}12{]}.

The stimulus fed to the system consists of $n$ components, each with
a Gaussian profile and a time-dependent fluctuation in strength. It
is given by 
\begin{eqnarray}
I_{0}^{\text{ext}}\left(x,t\right) & = & \sum_{i=1}^{n}\left[A_{0}+\delta A_{i}\left(t\right)\right]\exp\left(-\frac{\left|x-z_{i}\right|^{2}}{2a_{I}^{2}}\right).\label{eq:I0ext}
\end{eqnarray}
Here, $z_{i}$'s are the peak positions of the components, and $a_{I}$
is the width of the Gaussian profiles. If not specified, it was assumed
to be the same as the synaptic interaction range $a$ used in Eq.
(\ref{eq:Jxx}). $A_{0}$ is the average relative magnitude of one
input component, while $\delta A_{i}\left(t\right)$ is a random fluctuation
with standard deviation $\sigma_{A}$ in amplitude of input components.

\begin{figure}
\begin{centering}
\includegraphics[width=0.6\columnwidth]{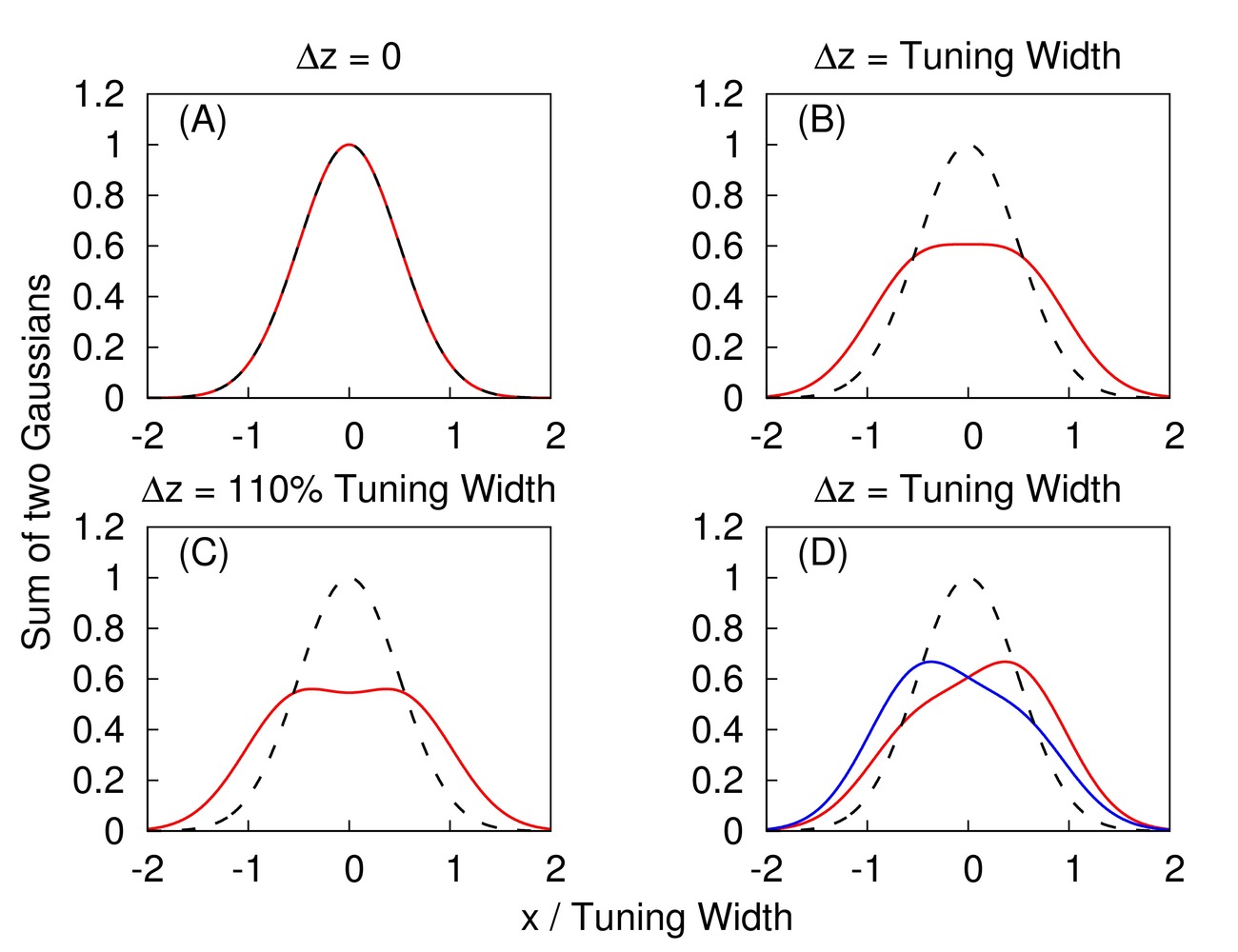} 
\par\end{centering}

\caption{\label{fig:two_gaussian} (A) - (C) The profile of two superposed
Gaussian functions with the same height. $f(x)\equiv\lbrace\exp[(x-\Delta z/2)^{2}/(2a^{2})]+\exp[(x+\Delta z/2)^{2}/(2a^{2})]\rbrace/2$.
Red solid line: $y=f(x)$ with different $\Delta z$. Dashed line:
$y=f(x)$ with $\Delta z=0$ as a reference. (A) $\Delta z=0$. (B)
$\Delta z$ = tuning width = $2a$. (C) $\Delta z$ = 110 \% tuning
width = $2.2a$. (D) The profile of two superposed Gaussian functions
with different heights to illustrate how the amplitude fluctuations
provide a cue to distinguish the components. $g(x)\equiv\lbrace A_{0}\exp[(x-\Delta z/2)^{2}/(2a^{2})]+A_{1}\exp[(x+\Delta z/2)^{2}/(2a^{2})]\rbrace$.
Dashed line: $y=f(x)$ with $\Delta z=0$ as a reference. Red solid
line: $y=g(x)$ with $\Delta z$ = tuning width, $A_{0}=0.4$ and
$A_{1}=0.6$. Blue solid line: $y=g(x)$ with $\Delta z$ = tuning
width, $A_{0}=0.6$ and $A_{1}=0.4$. }
\end{figure}

Note that when the Gaussian profiles have strong overlaps, the components
cannot be resolved, as illustrated in Figure \ref{fig:two_gaussian}(A)
- (C). We consider the amplitude fluctuations of each component to
be independent of each other, i.e, $\langle\delta A_{i}\delta A_{j}\rangle=0$,
where the average is over time. These fluctuations provide a cue for
the system to distinguish different components (Figure \ref{fig:two_gaussian}(D)).
This is consistent with the psychophysical experiment which showed
that spatial and temporal randomness is important for perception of
motion transparency \cite{Qian1994}. Since the fluctuations vanish
when averaged over time, a system responding only to time-averaged
inputs is unable be able to detect the components. Here, the role
of STD is to modulate the network state, so that it responds to one
input component once a time.

To model the situation that the maximum strength of the input profile
is invariant, we consider the input in Eq. (\ref{eq:dyn_u}) to be
\begin{equation}
I^{\text{ext}}\left(x,t\right)=\frac{A}{\max_{x}\left[I_{0}^{\text{ext}}\left(x,t\right)\right]}I_{0}^{\text{ext}}\left(x,t\right),\label{eq:Iext}
\end{equation}
where $A$ is the fixed maximum magnitude of the external input. As
the external input profile is set to have a constant maximum, only
the ratio $\sigma_{A}/A_{0}$, rather than the magnitudes of $A_{0}$
and $\sigma_{A}$, is relevant in our studies.

It is convenient to rescale the dynamical variables as follows. We
first consider the case without STD when $\beta=0$, and the synaptic
interaction range $a\ll L$. In this case, $p(x,t)=1$ in Eq. \eqref{eq:dyn_u}.
For $k\leq k_{c}\equiv\rho J_{0}^{2}/(8\sqrt{2\pi}a)$, the network
holds a continuous family of Gaussian-shaped stationary states when
$I^{\mathrm{ext}}(x,t)=0$. These stationary states are

\begin{eqnarray}
\tilde{u}(x) & = & \tilde{u}_{0}\exp\left(-\frac{\left|x-z\right|^{2}}{4a^{2}}\right)\text{, and}\\
\tilde{r}(x) & = & \tilde{r}_{0}\exp\left(-\frac{\left|x-z\right|^{2}}{2a^{2}}\right).
\end{eqnarray}
where $\widetilde{u}(x)$ is the rescaled variable $\varrho J_{0}u(x)$,
and $\tilde{u}_{0}$ is the rescaled bump height. The parameter $z$,
i.e., the center of the bump, is a free parameter, implying that the
stationary state of the network can be located anywhere in the space
$x$. In this paper, we assume that the variable is represented solely
by the peak position of the neural activity profile. This assumption
is one of the most direct ways to interpret the population code. However,
there are other ways to interpret population codes. For example, Treue
\etal. (2000) proposed that the curvature of the average of the neural
activity carries information represented by the neural population
code, although the mechanism achieving this objective is not clear
\cite{Treue2000}. On the phenomenological level, distributional population
coding and double distributional population coding were proposed to
represent information in population coding with more sophistication
\cite{Zemel2000,Sahani2003}.

The tuning width of a neuron, defined as the standard deviation of
the firing rate profile multiplied by 2, is therefore $2a$. In the
present work, we rescale the neuronal current as $\tilde{u}\left(x,t\right)\equiv\rho J_{0}u\left(x,t\right)$,
together with the corresponding rescaling of other variables given
by $\tilde{A}\equiv\rho J_{0}A$, $\tilde{k}\equiv k/k_{c}$, $\tilde{\beta}\equiv\tau_{d}\beta/(\rho^{2}J_{0}^{2})$.
By using these rescaling rules, the dynamics of the system should
only depend on $\tilde{k}$, $\tilde{\beta}$, $\tau_{d}/\tau_{s}$,
$\sigma_{A}/A_{0}$, $z_{i}$'s and $\tilde{A}$. Below, only these
parameters will be specified.

In each simulation, the variables $u(x,t)$ are modeled to be located
at $N$ discrete positions uniformly distributed in the space of preferred
stimuli $\left\{ x\right\} $. To do massive simulations, all simulation
results are generated by using $N=80$. We have verified that the
dynamics of the system is independent to $N$, and the number of neurons
should not affect the conclusion. The boundary condition of the space
is periodic. The range of the network is 360\textsuperscript{o} and
the tuning width of the neurons is 96\textsuperscript{o}, following
the experimental estimates in \cite{Treue2000}. To solve differential
equations in Eqs. (\ref{eq:dyn_u}) and (\ref{eq:dyn_p}), we used
the Runge-Kutta Prince-Dormand (8,9) method provided by the GNU Scientific
Library. Initial conditions of $u\left(x,t\right)$'s is zero, while
$p\left(x,t\right)$'s are initially 1. The local error of each evolution
step is less than 10\textsuperscript{-6}. The random number generator
used to generate the Gaussian random number is the generator proposed
by L\"{u}scher \etal. \cite{Luscher1994}. The Gaussian fluctuation
is updated every 50$\tau_{s}$.

\section{{\normalsize Results}}

\subsection{{\normalsize Population spikes}}

\begin{figure}
\begin{centering}
\includegraphics[width=1\textwidth]{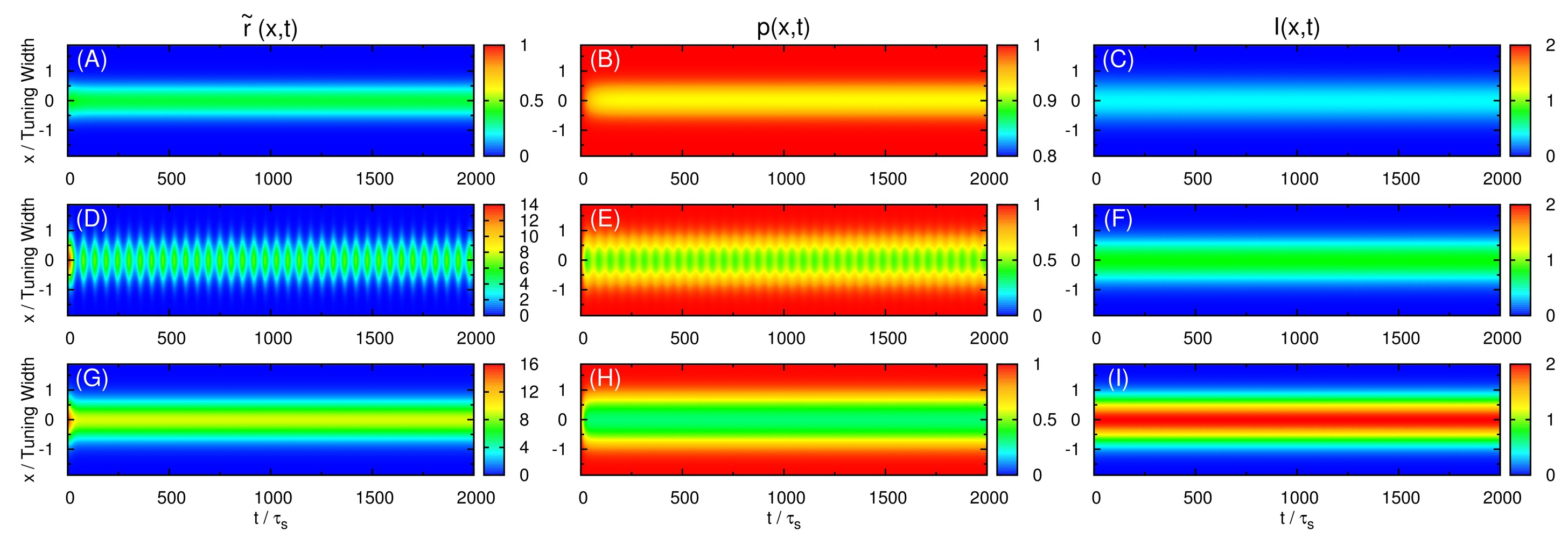} 
\par\end{centering}

\caption{\label{fig:population_spike}Firing rates $\tilde{r}\left(x,t\right)$
((A), (D) and (G)), available fraction of neurotransmitters $p\left(x,t\right)$
((B), (E) and (H)) and corresponding input (((C), (F) and (I))) for
various magnitudes of single-peaked external inputs. (A)-(C) $\tilde{A}=0.4$,
(D)-(F) $\tilde{A}=0.8$ and (G)-(I) $\tilde{A}=2.0$. Other parameters:
$\tilde{k}=0.5$, $\tilde{\beta}=0.24$, $a=48\pi/180$ and $\tau_{d}=50\tau_{s}$.}
\end{figure}

We first consider the response of the network when the input consists
of one component. We explore the network behavior by varying the parameters
$\tilde{k}$, $\tilde{\beta}$ and $\tilde{A}$. We found a rich spectrum
of behaviors including population spikes, static bumps, and moving
bumps. The full picture will be reported elsewhere. For the purpose
of the present paper, we fix $\tilde{k}$ and $\tilde{\beta}$ at
a typical value and consider the behavior when $\tilde{A}$ increases.
As shown in the top panel of Figure \ref{fig:population_spike}, the
network cannot be triggered to have significant activities when the
input is weak. In the bottom panel, the input is so strong that the
network response is stabilized to a static bump with time-independent
amplitude. An interesting case arises in the middle panel for moderately
strong input, where population spikes can be observed. Population
spikes are the consequence of the presence of STD. They are caused
by a rapid rise of neuronal activity due to the external stimulus.
Then in a time of the order of $\tau_{d}$, the neurotransmitters
are consumed, leading to a rapid drop in neuronal activity. When the
neurotransmitters recover, the neurons become ready for the next population
spike, resulting in the interesting periodic behavior. Population
spikes have been found before as synchronization of neuronal activities,
and their potential role in processing information was appreciated,
but no specific context of such applications was identified \cite{Loebel2002},
Here, we will present an example that spatially localized population
spikes endow the neural system a capacity of reading-out input components.

\subsection{{\normalsize Network activities for two stimuli}}

Next, we consider inputs with two components separated by $\Delta z>0$
and study the network behavior when $\Delta z$ gradually increases.
Without loss of generality, we choose $z_{1}=\Delta z/2$ and $z_{2}=-\Delta z/2$.
The relative fluctuation is $\sigma_{A}/A_{0}=0.3$.

\begin{figure}
\begin{centering}
\includegraphics[width=0.6\columnwidth]{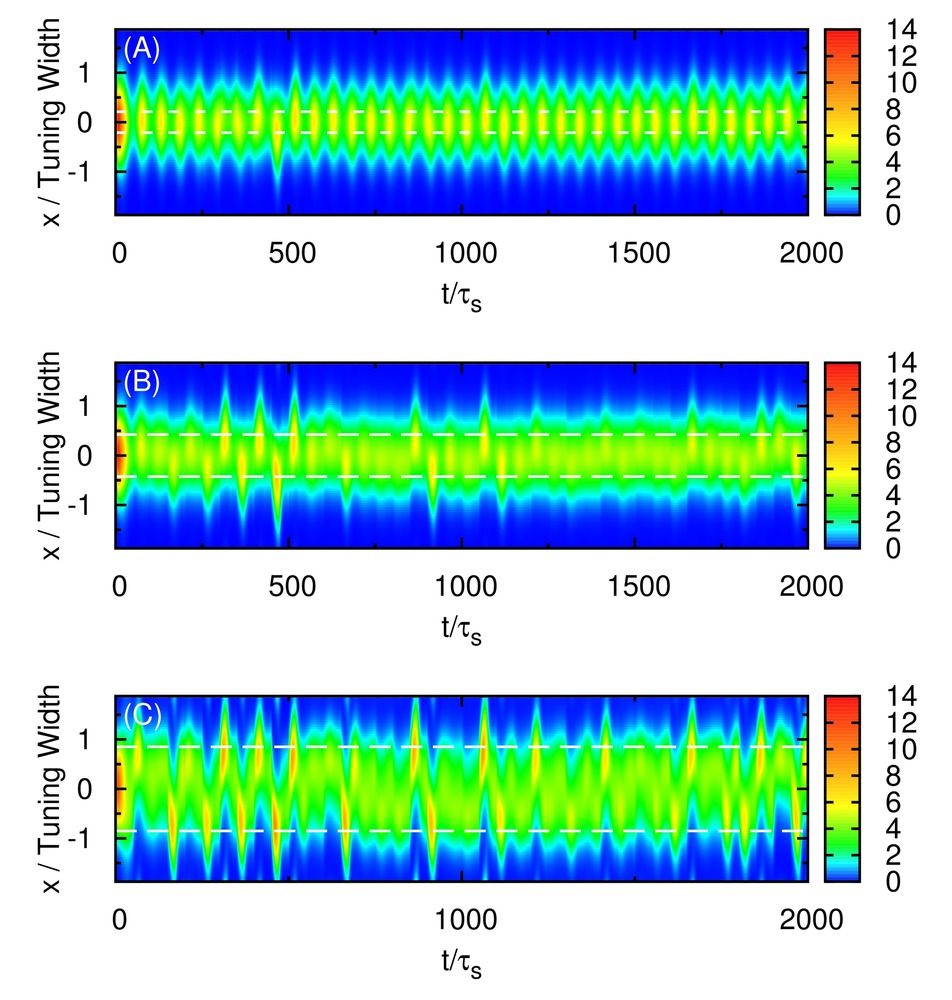} 
\par\end{centering}

\caption{\label{fig:network_act_var_dz}Raster plot of firing rates $\tilde{r}$
for (A) $\Delta z=0.5$, (B) $\Delta z=1.0$ and (C) $\Delta z=2.0$.
White dashed lines: positions of stimuli. Parameters: Other parameters:
$\tilde{k}=0.5$, $\tilde{\beta}=0.24$, $a=48\pi/180$, $\tilde{A}=0.8$,
$\sigma_{\delta A_{i}}/A_{0}=0.3$ and $\tau_{d}=50\tau_{s}$.}
\end{figure}

When the separation is small, the positions of the population spikes
fluctuate around the mid-position of the two stimuli, as illustrated
in Figure \ref{fig:network_act_var_dz}(A). The two components cannot
be resolved.

When the separation increases to the extent that the two components
remain barely resolved, an interesting change in the spiking pattern
occurs as shown in Figure \ref{fig:network_act_var_dz}(B). The positions
of the population spike peaks begin to center around the two input
components, although the shoulders of the population spikes remain
overlapping considerably. Note that in this regime, the profile of
the neuronal activities remain unresolved when they are averaged over
time. However, due to the presence of STD, it is likely that a population
spike is produced at the position of the component which happens to
be higher due to height fluctuations. Hence in this regime, the population
spike peaks are no longer aligned at the center. Rather, they are
arranged in two rows, each around the two components. Furthermore,
the two rows of population spikes tend to fire alternately. This implies
that although it is hard to resolve the two components by considering
the time-averaged signals, the temporal modulation by the alternating
population spikes may be utilized for resolution enhancement.

When the separation increases further, the population spikes form
two groups clearly, as shown in Figure \ref{fig:network_act_var_dz}(C).
The two components are clearly resolved.

\begin{figure}
\begin{centering}
\includegraphics[width=1\textwidth]{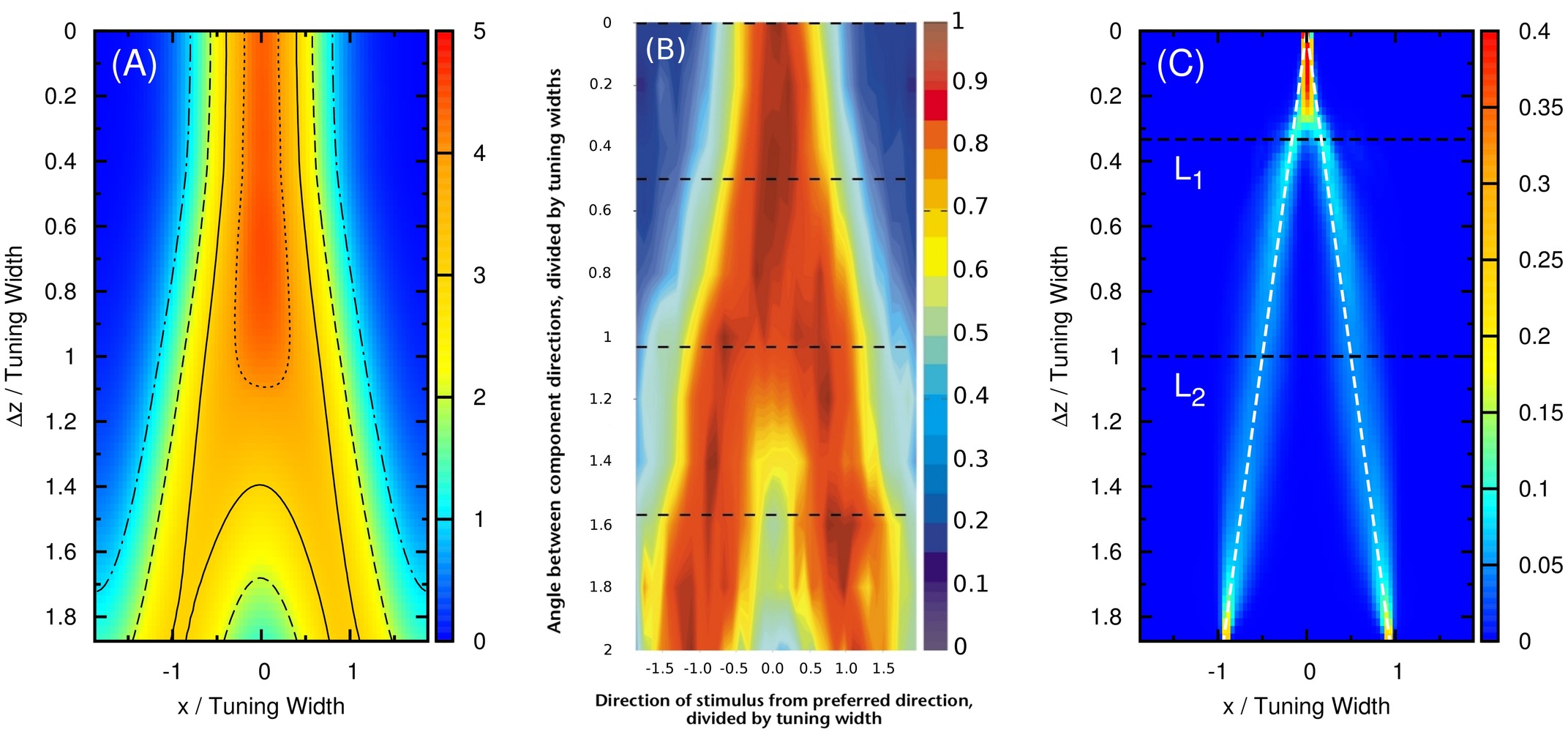} 
\par\end{centering}

\caption{\label{fig:average_activity}(A) Time average of firing rates $\tilde{r}$
as a function of the preferred stimuli of neurons, $x$, and the separation
between the two stimuli, $\Delta z$. Contour lines: $\left\langle \tilde{r}\right\rangle _{t}=1$
(dotted-dashed line), $\left\langle \tilde{r}\right\rangle _{t}=2$
(dashed line), $\left\langle \tilde{r}\right\rangle _{t}=3$ (solid
line), $\left\langle \tilde{r}\right\rangle _{t}=4$ (dotted line).
Parameters: same as Figure \ref{fig:network_act_var_dz}. (B) The
average neural activity recorded by Treue \etal in \cite{Treue2000}
(with license number 3125800919243 for the reuse purpose). (C) Contours
of the distribution of peak positions higher than 6.2 as a function
of preferred stimuli, $x$, and the separation between the two stimuli,
$\Delta z$. White dashed line: positions of the two stimuli. $L_{1}$:
one-third of the tuning width. $L_{2}$: tuning width. Parameters:
same as Figure \ref{fig:network_act_var_dz}.}
\end{figure}

To compare our model with experimental results, we measure the time
average of neuronal activities as a function of preferred stimuli
of neurons and the separation of the two stimuli, shown in Figure
\ref{fig:average_activity}(A). We found that this result is very
similar to the experimental results reported by Treue \etal. (Figure
2(C) in \cite{Treue2000}). The peak of the average profile of neuronal
activities splits near $\Delta z\sim1.0\times$tuning width. However,
the time-averaged data cannot explain why subjects can resolve separations
much less than the tuning width.

\subsection{{\normalsize Extraction of Modulated Information}}

\begin{figure}
\begin{centering}
\includegraphics[width=0.6\columnwidth]{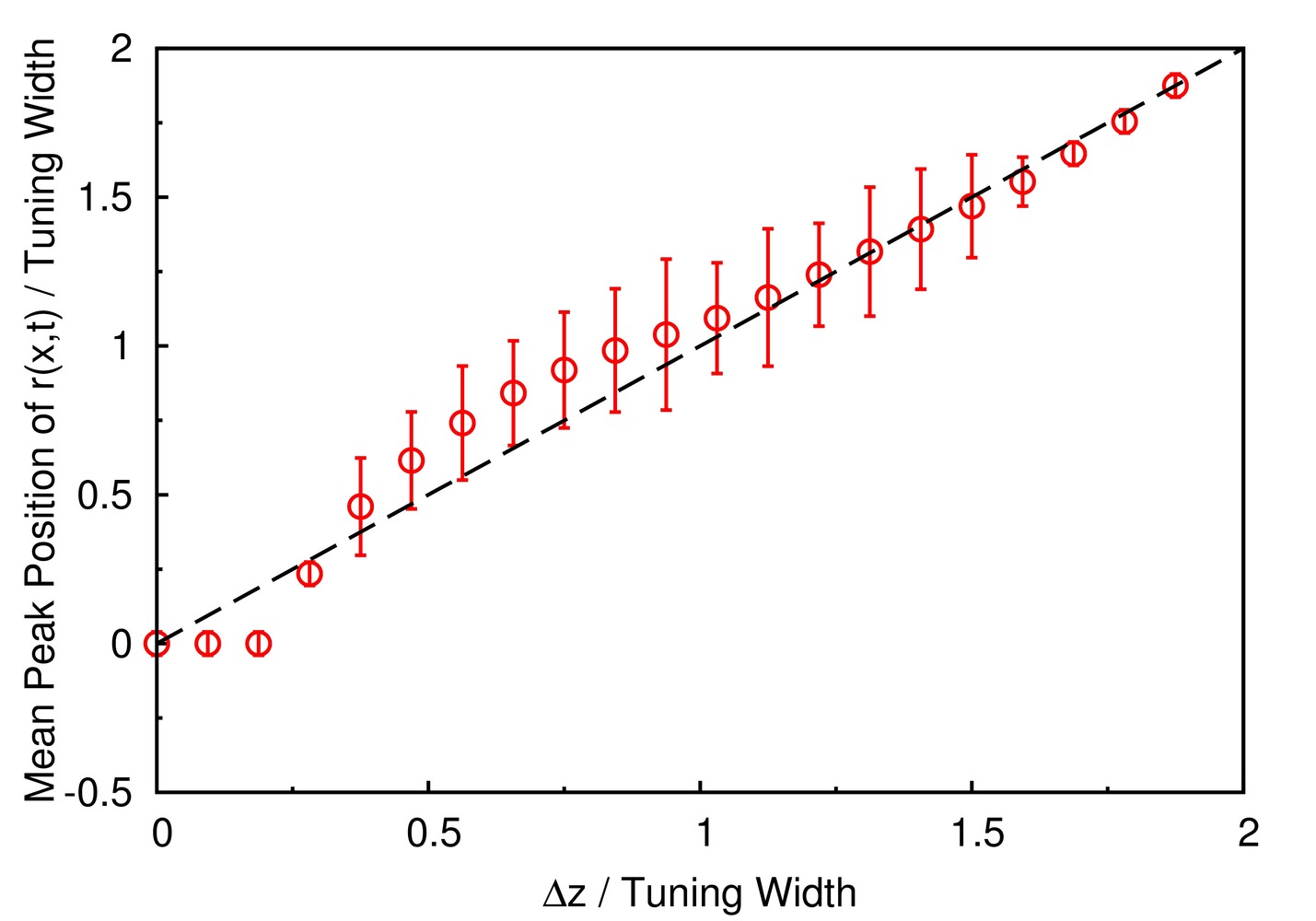}
\par\end{centering}

\caption{\label{fig:mean_sep} The mean separation of peak positions of $r(x,t)$
shown in Figure \ref{fig:average_activity}(C). Symbols: simulation.
Dashed line: diagonal line representing perfect distinguishability.}
\end{figure}

To demonstrate that the neuronal activities carry the information
about two stimuli, we collect statistics on the peak positions of
the population spikes. Here the peak position is calculated by $\max{}_{x}\tilde{r}(x)$.
In Figure \ref{fig:average_activity}(C), we present the contour plot
of the distribution of peak positions in the space of the preferred
stimuli of neurons and separation between the two stimuli in units
of the tuning width. To focus on peaks with significant information
only, we counted only population spikes with maximum amplitudes above
an appropriately chosen threshold. Each column in Figure \ref{fig:average_activity}(C)
is a normalized histogram with 80 bins. In order to obtain a relatively
smooth distribution, the sampling process lasted for 100000$\tau_{s}$.
The mean of the separation between peak positions is plotted in Figure
\ref{fig:mean_sep} as a function of $\Delta z$. We found that in
this setting, the system can detect the input separation down to one-fourth
of the tuning width. We note that in Figure \ref{fig:average_activity}(C),
when the difference between the components is too small, $\Delta z\lesssim\nicefrac{1}{4}\text{ tuning width}$,
population spikes occur at the middle of the net external input profile
with a relatively small variance. However, when the network starts
to resolve the two components, there are notable variances on positions
of the population spikes in each component. The standard deviation
of the positions of the population spikes in each component is roughly
of the order of 0.1 times the tuning width, which is roughly 20\textdegree{},
as shown in Figure \ref{fig:mean_sep}.

To investigate whether the statistics with long sampling period is
applicable to sampling periods in actual experiments, we have also
collected statistics for $500\tau_{s}$. (In the experiment done by
Treue \etal., subjects took 500 ms to perform the discriminational
task.) The result is shown in Figure \ref{fig:fast_decode}(A) in
Appendix. Although the distribution is rougher because of the relatively
small sampling size, enhanced resolution down to 0.3 tuning width
is still visible.

Furthermore, when the separation between the two stimuli lies between
one-third and three-halves of the tuning width, the system slightly
overestimated the separation of the two profiles. If we take the tuning
width to be 96\textdegree{} \cite{Treue2000}, this range will be
approximately from 30\textdegree{} to 140\textdegree{}. This is consistent
with the experimental results of Braddick \etal. \cite{Braddick2002},
in which subjects overestimated some moving direction difference in
transparent motion experiments. However, it was reported in Fig. 4
in \cite{Treue2000} that the perceived separation of movement direction
starts to underestimate the truth when the stimulus separation increases
above 40º. Since the range corresponding to `motion repulsion' reported
by Braddick \etal. \cite{Braddick2002}. is different from that reported
by Treue \etal., it seems that the range of differences between stimuli
corresponding to `motion repulsion' is different for different experimental
settings.

We have also tested the effects of choosing the widths of external
input components to be different from the tuning width of the neuronal
response. We found that the results for different stimulus strengths
in Figure \ref{fig:prob_den_cutoff_var_aI} in Appendix are qualitatively
the same as that in Figure \ref{fig:average_activity}(C). 

\begin{figure}
\begin{centering}
\includegraphics[width=0.8\textwidth]{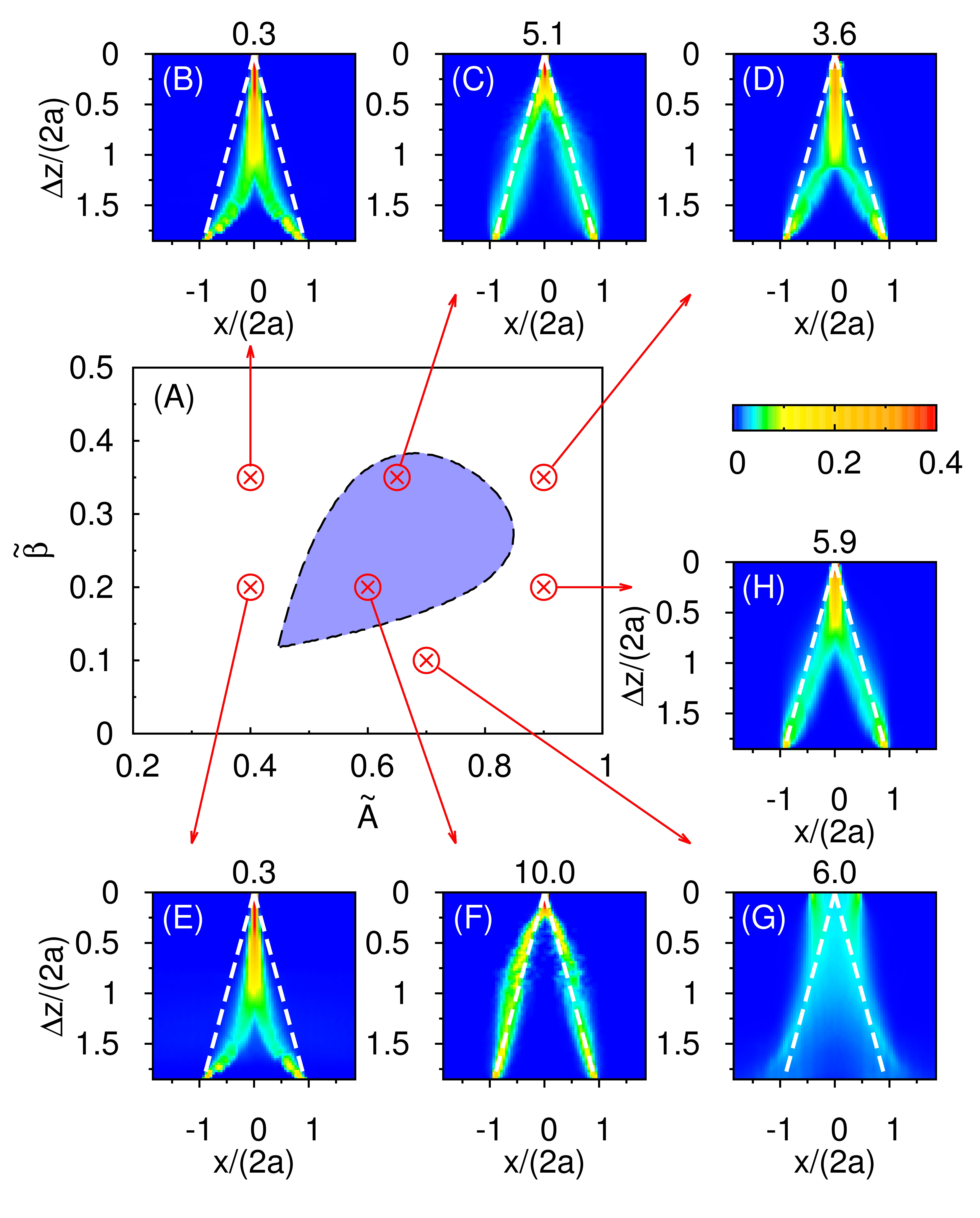}
\par\end{centering}

\caption{\label{fig:phase_diagram}(A) The phase diagram of population spikes
over the parameter space spanned by $(\tilde{A},\tilde{\beta})$ with
the parameter $\tilde{k}=0.5$ and $\tau_{d}/\tau_{s}=50$. (B) -
(H) are distributions of the occurence of peak positions as function
of $\Delta z$. The numbers at the top of (B) - (H) are thresholds
used to sample peak positions. Parameters: (B) $\tilde{A}=0.4$ and
$\tilde{\beta}=0.35$. (C) $\tilde{A}=0.65$ and $\tilde{\beta}=0.35$.
(D) $\tilde{A}=0.9$ and $\tilde{\beta}=0.35$. (E) $\tilde{A}=0.4$
and $\tilde{\beta}=0.2$. (F) $\tilde{A}=0.6$ and $\tilde{\beta}=0.2$.
(G) $\tilde{A}=0.7$ and $\tilde{\beta}=0.1$. (H) $\tilde{A}=0.9$
and $\tilde{\beta}=0.2$.}
\end{figure}

The result shown in Figure \ref{fig:average_activity}(C) is not particular
for the chosen set of parameters. In Figure \ref{fig:phase_diagram},
there is a phase diagram along with some selected parameters. In Figure
\ref{fig:phase_diagram}(A), the colored region is the region for
population spikes with one stimulus. If $\tilde{A}$ and $\tilde{\beta}$
are chosen from this region, as far as we have observed, similar results
can be obtained by choosing appropriate thresholds. If $\tilde{A}$
and $\tilde{\beta}$ are outside the colored region, no matter what
the threshold was, the result shown in Figure \ref{fig:average_activity}(C)
cannot be reproduced. This result suggests that population spikes
are important to resolution enhancement.

\subsection{{\normalsize Network response with multiple stimuli}}

\begin{figure}
\begin{centering}
\includegraphics[width=0.55\textwidth]{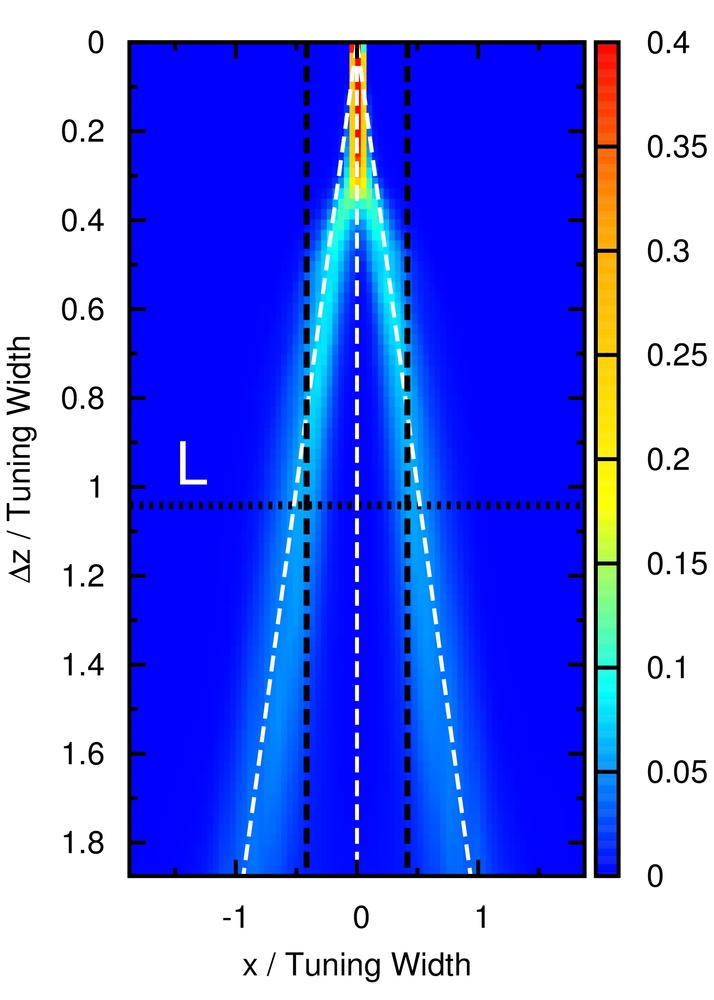} 
\par\end{centering}

\caption{\label{fig:prob_den_cutoff_3sti} Contours of the distribution of
peak positions \textcolor{black}{higher than 6.2} as a function of
preferred stimuli, $x$, and the separation between the two outermost
stimuli, $\Delta z$, in the case of three equally strong stimuli.
White dashed line: positions of three stimuli. Horizontal dotted line:
the case comparable to the three-stimulus experiment reported by Treue
\etal. Vertical dashed lines: perception ($\pm40$\textdegree{})
reported by subjects in the experiment in units of the tuning width
($96$\textdegree{}). Parameters: same as Figure \ref{fig:mean_sep}.}
\end{figure}

We further test the response of our model to more than two stimuli.
Figure \ref{fig:prob_den_cutoff_3sti} shows the case for three stimuli
of equal amplitude, whose peak positions are labeled by the white
dashed lines. However, the contours of the distribution of population
spikes are double-peaked, similar to those in Figure \ref{fig:mean_sep}.
This result suggests that, if there are three stimuli overlapped together,
the network response should give only two groups of neuronal responses.
Also, it predicts that peaks of population spikes should occur at
positions that underestimate the separation between the outermost
stimuli. A similar result for shorter sampling periods comparable
to actual experiments can be found in Figure \ref{fig:fast_decode}(B)
in Appendix.

We found that the experimental result of multiple stimuli reported
by Treue \etal. is consistent with this prediction. In their paper,
it was reported that, when there were three groups of moving dots
moving at directions \textpm{}50\textdegree{} and 0º, the subjects
would report that there were only two moving directions at \textpm{}40\textdegree{}.
This consistency is shown in Figure \ref{fig:prob_den_cutoff_3sti},
where the vertical dotted line $L$ labels the position that the outermost
stimuli are directed at \textpm{}50\textdegree{} when the tuning width
is 96\textdegree{}, and the pair of horizontal dashed lines labels
\textpm{}40\textdegree{} correspondingly.

\section{{\normalsize Conditions for resolution enhancement}}

We have demonstrated the phenomenon of resolution enhancement due
to modulations of population spikes. To see whether this picture can
be generalized to other cases and what alternative models are to be
excluded, we summarize the general conditions of its occurrence. To
appreciate the significance of each condition, we will consider the
alternative scenarios in the presence and absence of the various conditions.

\subsection{Short-term synaptic depression}

Without the STD, the steady state of the neuronal activity profile
becomes centered at either one of the two input stimuli. In Figure
\ref{fig:conditions}(A), when the difference between the input profiles
is large, $\Delta z/a=3.7$ for instance, the neuronal activity is
trapped by the input profile near $x=1.55$. This case is not consistent
with experiments, because when the separation between the input profiles
is large enough, the neuronal activity should be able to identify
both stimuli. This shows that STD plays the following roles in this
phenomenon.

First, STD gives rise to the temporal modulation characterized by
the population spikes, in which rapid rises in population activities
alternate periodically with drops due to the consumption of neurotransmitters.
Spiking activities enable the activity profile to jump from one stimulus
position to another easily.

Second, the presence of STD enhances the mobility of the activity
profiles. Due to the consumption of neurotransmitters in the active
region, the profile tends to relocate itself to less active regions.
This is the cause of the increased mobility when the activity profile
tracks the movement of external stimuli, as well as their anticipatory
tracking as a possible mechanism for delay compensation \cite{Fung2012a,Fung2012b}.
In the parameter regime where the stationary profile becomes unstable
in its position, and population spikes become the attractor state,
the network tends to establish a population spike in new locations,
preventing itself from being trapped by one stimulus. This results
in population spikes centered at alternating stimuli and hence the
temporal modulation.

For example, if the two stimuli are strongly overlapped, the average
neuronal response concentrates at the in-between region of the two
stimuli, as shown in Figure \ref{fig:network_act_var_dz}(B). In this
case, the time-average profile of the dynamical variable $p\left(x,t\right)$
has a dip centered at the midpoint between two stimuli, as shown in
Figure \ref{fig:average_p_1.0tw}. Since, in our model, there are
fluctuations of the magnitude of each component of the external input,
population spikes occur near the positions of the stimuli, labeled
by the blue lines in Figure \ref{fig:average_p_1.0tw}. Since the
synaptic efficacies of the presynaptic neurons are stronger in the
side region further away from the other stimulus, population spikes
are more likely to happen in the outer region rather than the inner
region. So, the separation between the two groups of population spikes
can be larger than the separation between the two stimuli. This is
also the reason why only two groups of population spikes can be observed
in the case with three stimuli (Figure \ref{fig:prob_den_cutoff_3sti}).
STD also explains the slight over-estimation of the perceived positions
when the separation of the stimuli is around the tuning width.

Third, when STD is not sufficiently strong, we observe that sloshers
rather than population spikes are formed \cite{Folias2011}. These
sloshers are bumps that oscillate back and forth around the external
stimuli, as shown in \textcolor{black}{Figure}\textcolor{red}{{} }\ref{fig:slosher}.
The height of the bumps is highest when they slosh to the extreme
positions, but due to the weaker STD, the height variation in a cycle
is not as extreme as those in the population spikes. The positional
extent of their oscillations is mainly determined by the restoring
attraction from the external input, and is effectively insensitive
to the stimulus profile. Hence in the task of resolving the stimulus
directions, the performance is degraded by the very flat part of the
curve of the perceived separation when the stimuli have strong overlaps,
as shown in Figure \ref{fig:phase_diagram}(G).

There are also other variants of the model that demonstrate the significance
of STD in similar ways. For example, in recurrent networks with local
inhibition, we may replace $B\left(t\right)$ in Eq. \eqref{eq:rxt}
by $B^{\prime}\left(x,t\right)$ given by 
\begin{equation}
B^{\prime}\left(x,t\right)=1+\rho k\int dx^{\prime}\exp\left(-\frac{\left|x-x^{\prime}\right|^{2}}{2b^{2}}\right)u\left(x^{\prime},t\right)^{2}.
\end{equation}
To stabilize the neural activity, the range of the local inhibition,
$b$, has to be larger than the range of excitatory connection, $a$.
However, if $a$ is as large as 48º, this local inhibtion can be fairly
replaced by $B\left(t\right)$ with appropriate $\tilde{k}$. In the
presence of STD, the discrimination performance is comparable to that
in Figure \ref{fig:mean_sep}, but but the resolution is poor otherwise.

\subsection{Suitably strong input profiles}

Suitably strong input magnitude is needed to produce the temporally
modulated patterns, as illustrated in Figure \ref{fig:phase_diagram}.
First, when the magnitude of the external input is too small, no significant
system-driven neuronal activity can be observed. Fluctuations of external
input components cannot stimulate the population spike, as the activation
by input profiles was not strong enough. Second, even when the magnitude
of the external input is larger, population spikes can be produced
but the stimulus is too weak to pin them at the position of the stimuli.
Since the mobility of the population spikes is enhanced by STD, moving
population spikes are formed, as illustrated in Figure \ref{fig:conditions}(B).
Since the population spikes move away from the stimulus positions
after their formation, they cannot be used to encode the stimulus
positions and also become part of the noisy background affecting the
recognition of the stimulus positions. When the stimulus is too strong,
population spikes cannot be generated and the resolution degrades.

\subsection{Fluctuations in input profiles}

Fluctuations on external input components is important to the behavior
in Figure \ref{fig:network_act_var_dz}. If there were no fluctuations
in the input profiles, the net input profile will have only one peak
for $\Delta z<2a$. As a result, there is effectively one bell-shaped
input profile if the difference between two stimuli is too small,
and the network response will also be single-peaked, as shown in Figure
\ref{fig:conditions}(C). Hence fluctuations in the external input
play the role of rendering the components distinguishable. As shown
in Figure \ref{fig:fluctuation}, recognition of input location always
follows a strong input on the same side at the current step, and a
strong input on the other side in the previous step, suggesting that
a sudden shift in input bias provides condition for reliable recognition.
In fact, the noise fluctuations act as the signals themselves, without
which the single-peaked input provides little information about the
components. Results in Figure \ref{fig:fluctuation} also illustrate
that, statistically, the system is able to give valid responses to
stimulus changes in a single step. This explains why the network yields
discrimination performance equally well for short and long sampling
periods, as demonstrated in a comparison between Figs. \ref{fig:average_activity}(C)
and \ref{fig:fast_decode}(A).

The fluctuations may come from randomness in the inputs. Psychophysical
experiments show that spatial and temporal randomness is important
for perceptions of motion transparency. For example, regularly spaced
lines moving in opposite directions do not give the perception of
transparent motion, whereas randomly spaced lines are able to do so
\cite{Qian1994}. The input signals come from different locations
of the visual field, and fluctuations arise when the perceived objects
move from one location to another. Fluctuations may also arise when
feedback signals from advanced stages of processing guide the system
to shift its attention from one specific component to another.

Functionally, fluctuations facilitate the resolution of the directional
inputs in the following two aspects. Spatially, it breaks the symmetry
of the input profile. Temporally, it provides the time-dependent signals
that induce the population spikes centered at the component that happens
to be strengthened by fluctuations. This enables the system to recognize
the temporally modulated inputs. On the other hand, for systems processing
only time-averaged inputs, the height fluctuations vanish when averaged
over time, so that the components cannot be detected.

\subsection{Thresholding}

\begin{figure}
\begin{centering}
\includegraphics[width=0.8\textwidth]{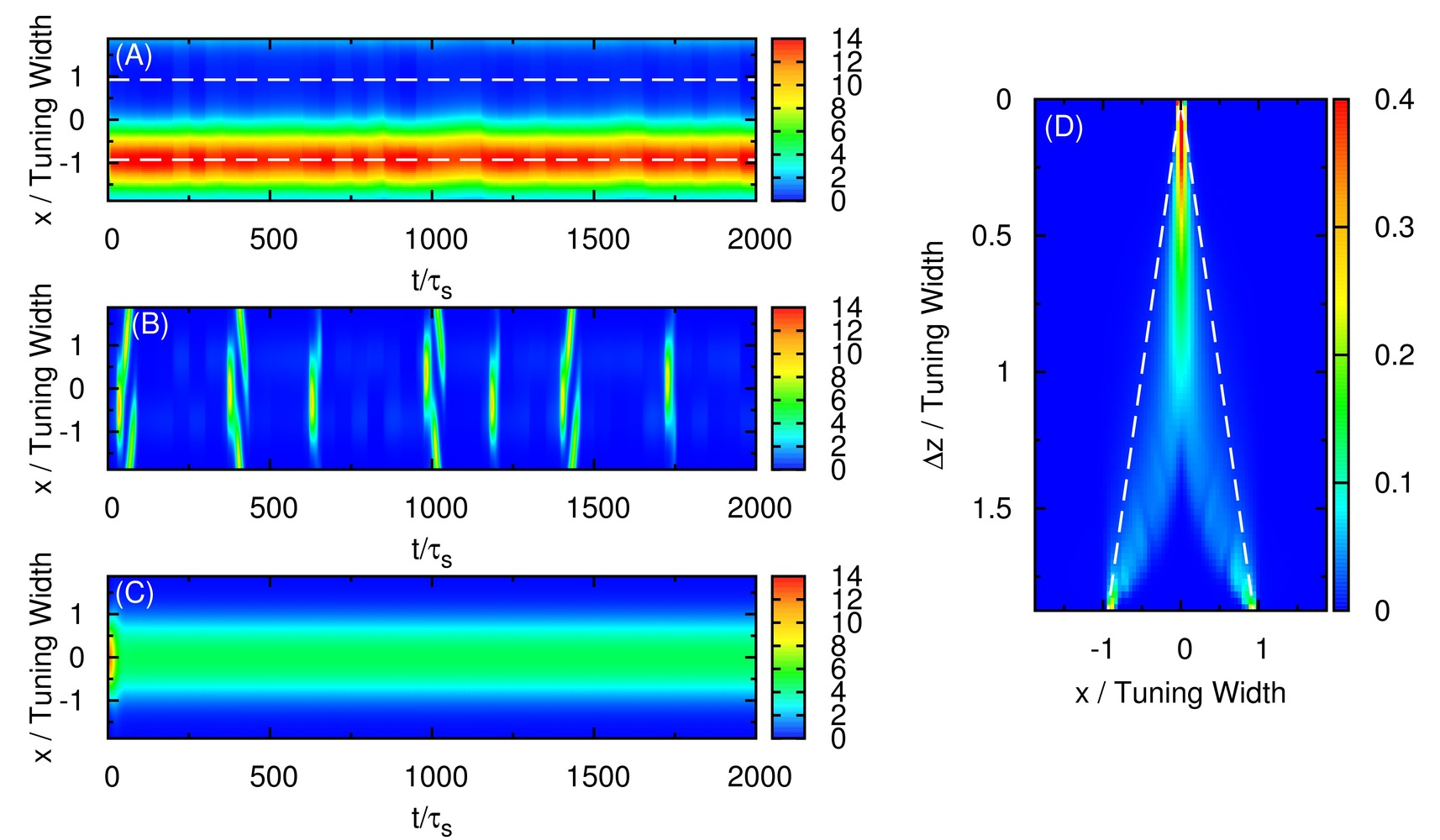} 
\par\end{centering}

\caption{\label{fig:conditions}(A) Raster plot of firing rate $\widetilde{r}$
of the network with two stimuli and without STD. Parameters: $\tilde{k}=0.5$,
$\tilde{\beta}=0$, $\tilde{A}=0.8$, $a=48\pi/180$, $\sigma_{A}/A_{0}=0.3$
and $\Delta z=3.1$. (B) Rastor plot of firing rate $\widetilde{r}$
of the network with two stimuli with weak net input profile. Parameters:
$\tilde{k}=0.5$, $\tilde{\beta}=0.24$, $\tilde{A}=0.4$, $a=48\pi/180$,
$\sigma_{A}/A_{0}=0.3$ and $\Delta z=2.5$. (C) Rastor plot of firing
rate $\widetilde{r}$ of the network with two stimuli without height
fluctuations in the external input profile. Parameters: $\tilde{k}=0.5$,
$\tilde{\beta}=0.24$, $\tilde{A}=0.8$, $a=48\pi/180$, $\sigma_{A}/A_{0}=0$
and $\Delta z=1.67$. (D) Contours of the distribution of peak positions
for all peak heights. White dashed line: positions of the two stimulus
components. Parameters: $\tilde{k}=0.5$, $\tilde{\beta}=0.24$, $\tilde{A}=0.8$,
$a=48\pi/180$, $\sigma_{A}/A_{0}=0.3$ and $\Delta z=1.0$.}
\end{figure}

\begin{figure}
\begin{centering}
\includegraphics[width=0.6\columnwidth]{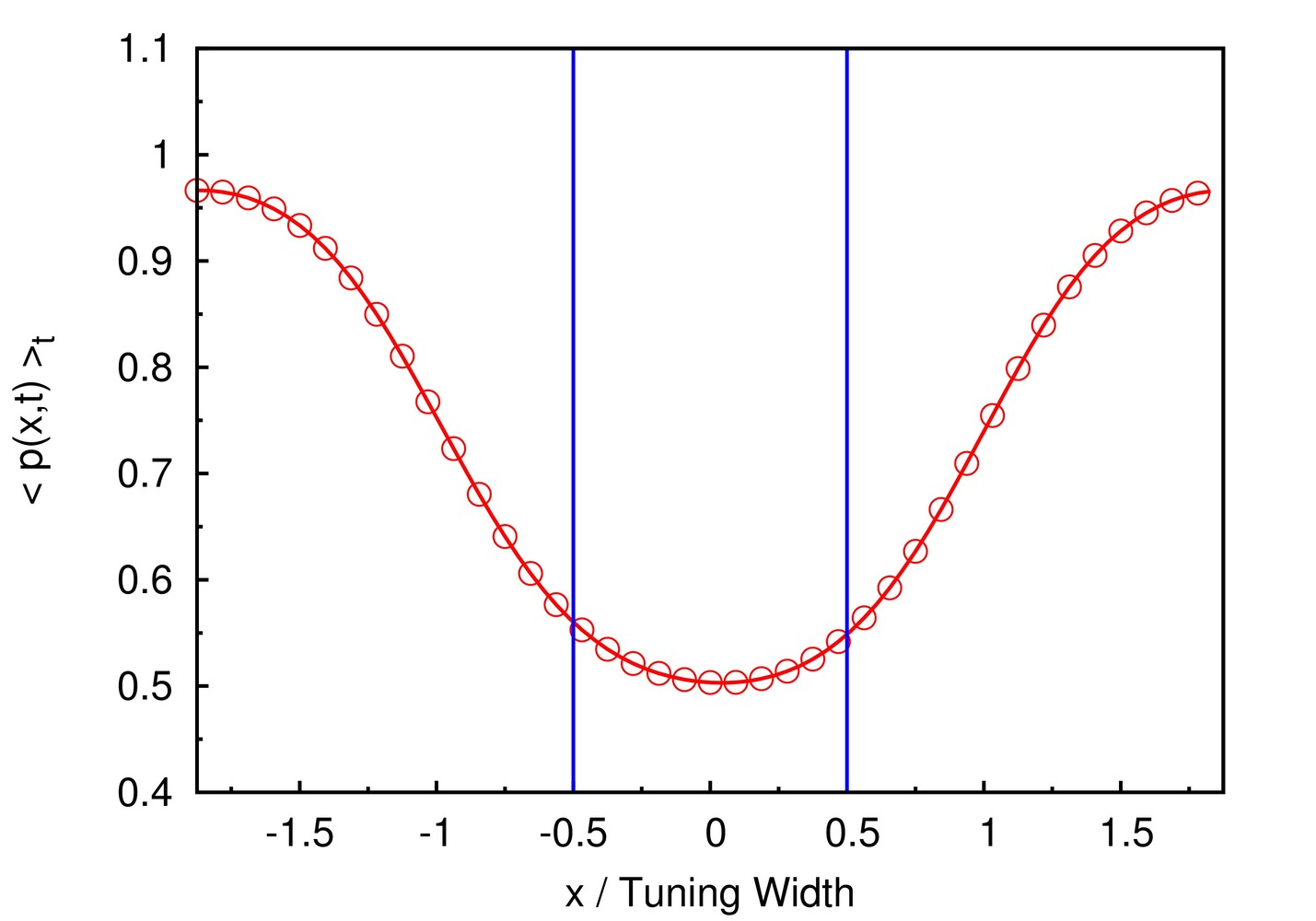} 
\par\end{centering}

\caption{\label{fig:average_p_1.0tw}The time-averaged dynamical variable $p\left(x,t\right)$.
Symbols and red line: measurement from the simulation. Blue lines:
positions of two stimuli. Parameters: $\tilde{k}=0.5$, $\tilde{\beta}=0.24$,
$a=48\pi/180$, $\tau_{d}/\tau_{s}=50$, $\tilde{A}=0.8$ and $\Delta z=$tuning
width of attractor states.}
\end{figure}

\begin{figure}
\begin{centering}
\includegraphics[width=0.85\textwidth]{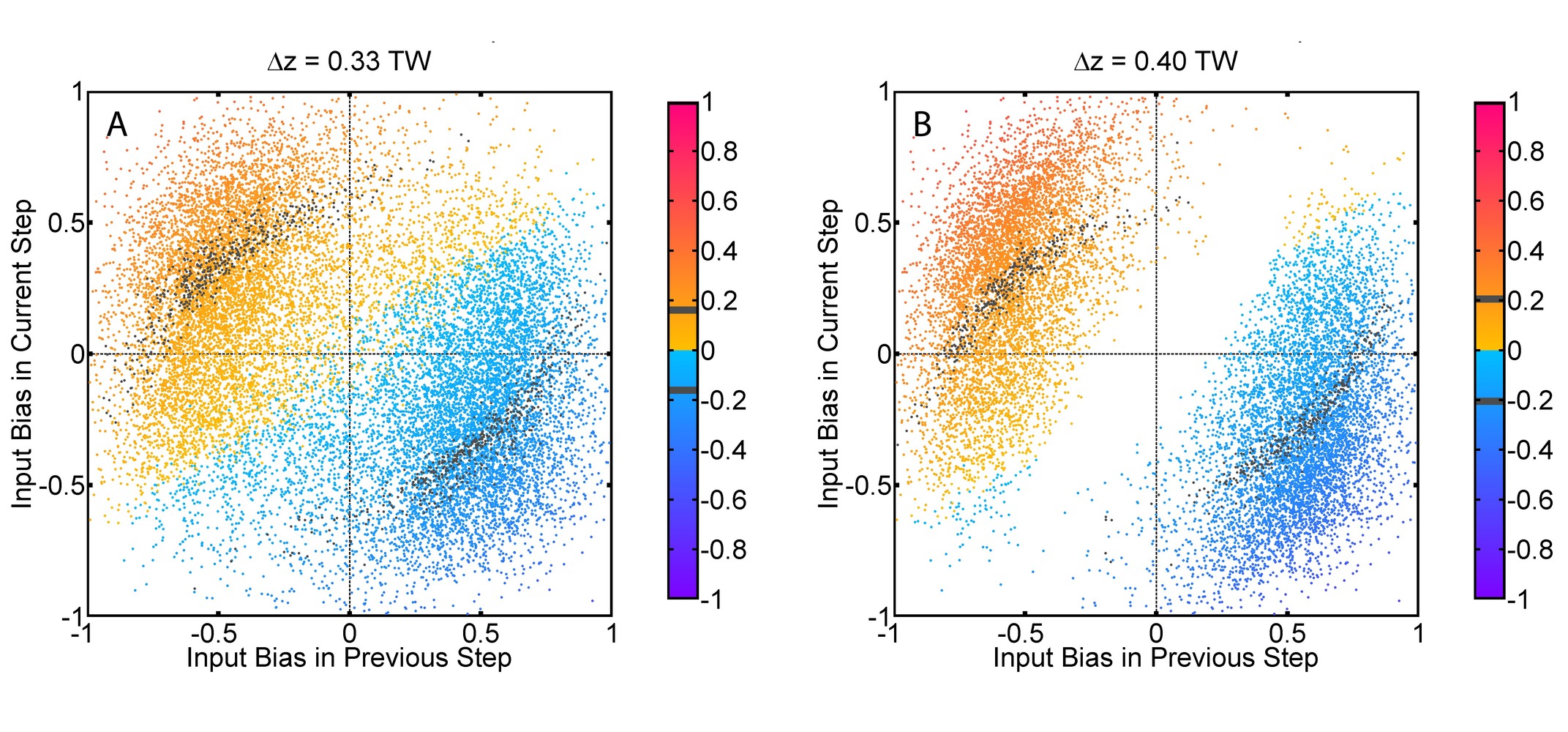}
\par\end{centering}

\caption{\label{fig:fluctuation}Population spikes' positions conditional on
input fluctuations. One step refers to $50\tau_{s}$, which is the
temporal interval between every update in the Gaussian fluctuation
$\delta A_{i}\left(t\right)$. Input bias is defined as $(\delta A_{1}\left(t\right)-\delta A_{2}\left(t\right))/\max(A_{0}+\delta A_{1}\left(t\right),A_{0}+\delta A_{2}\left(t\right))$.
The color code indicates the average position of population spikes
above threshold within one step in unit of the tuning width (TW).
Gray color means the average position is within the true position
of either input $\pm0.01$ TW. True positions of inputs:$z_{1}=\Delta z/2$,
$z_{2}=-\Delta z/2$. (A) $\Delta z=0.33$ TW. (B) $\Delta z=0.40$
TW. Parameters: same as Figure \ref{fig:network_act_var_dz}.}
\end{figure}

\begin{figure}
\begin{centering}
\includegraphics[width=0.6\columnwidth]{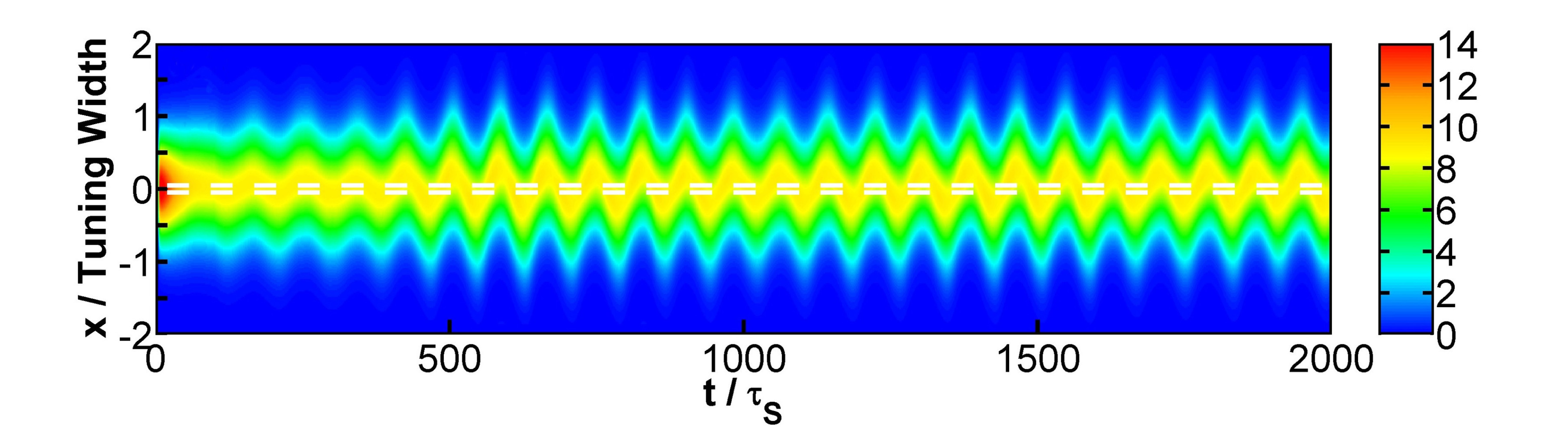} 
\par\end{centering}

\caption{\label{fig:slosher}Raster plot of firing rates $\tilde{r}$ at $\Delta z=0.1$,
showing a slosher. White dashed lines: positions of the stimuli. Other
parameters: $\tilde{k}=0.5$, $\tilde{\beta}=0.1$, $a=48\pi/180$,
$\tilde{A}=0.8$, $\sigma_{A}/A_{0}=0.2$ and $\tau_{d}=50\tau_{s}$.}
\end{figure}

Even after temporal modulation, resolution based on the network response
can still carry large errors. As shown in Figure \ref{fig:network_act_var_dz}(C),
there are obviously two groups centering around the positions of the
two components, but in between the two components, there is a region
with moderate neuronal activities. If the network includes neuronal
activities of all magnitudes, the errors in estimating the component
positions will be large, especially when $\Delta z$ is small. Indeed,
Figure \ref{fig:conditions}(D) shows that without imposing any thresholds
on the neuronal activities, the network cannot resolve the two components
until the separation exceeds the tuning width.

\begin{figure}
\begin{centering}
\includegraphics[width=0.8\columnwidth]{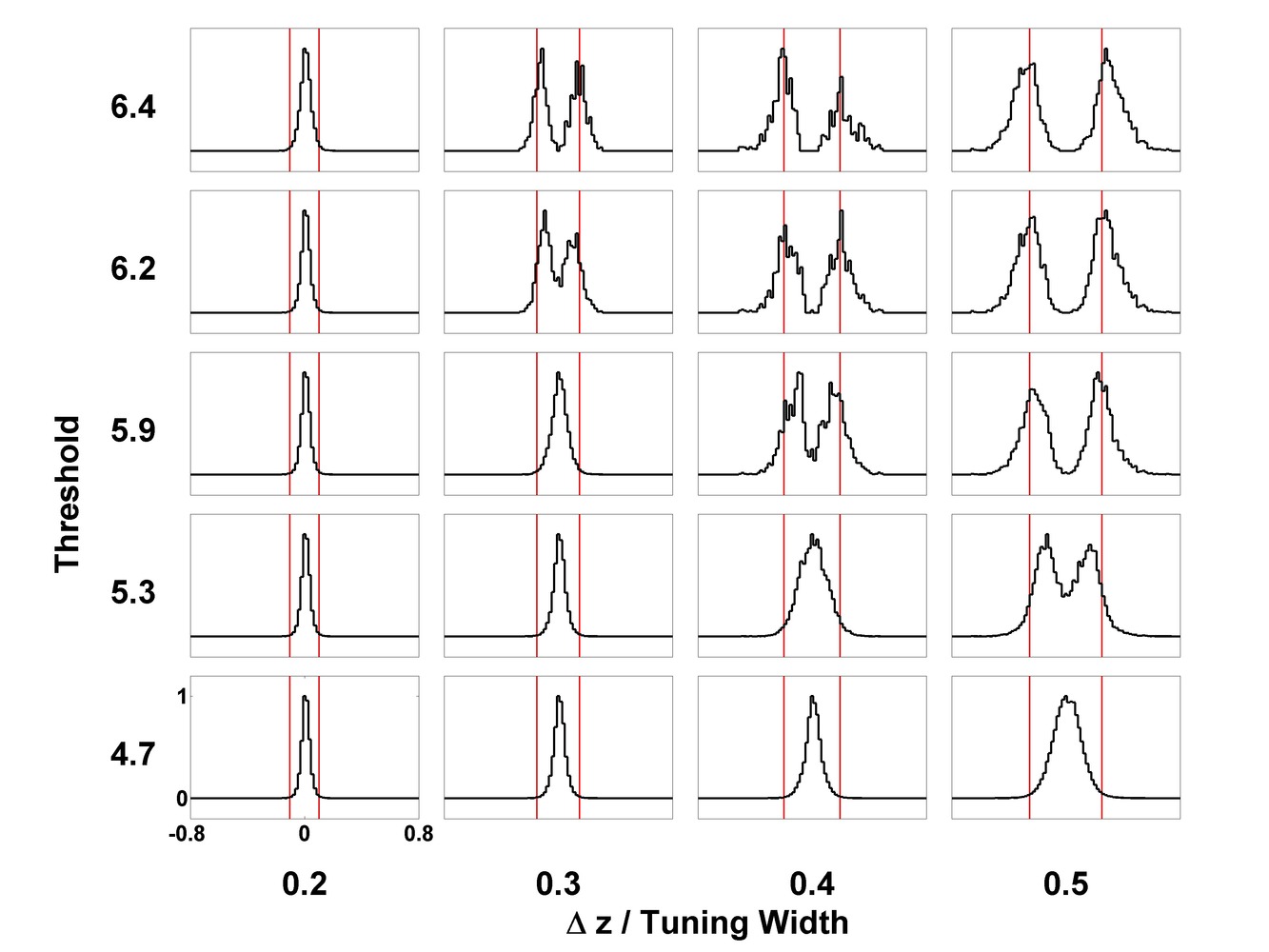} 
\par\end{centering}

\caption{\label{fig:Effect-of-threholding}Effect of thresholding on the statistics
of peak positions. Distributions of peak positions higher than different
thresholds (shown on the left) when two stimuli are separated by $\Delta z$
(shown at the bottom) are plotted in patches. The scales of all the
patches are the same, shown on the lower-left patch. Distributions
are normalized to the maximum value. Red bars mark the positions of
two stimuli. Other parameters: $\tilde{k}=0.5$, $\tilde{\beta}=0.1$,
$a=48\pi/180$, $\tilde{A}=0.8$, $\sigma_{A}/A_{0}=0.2$ and $\tau_{d}=50\tau_{s}$.}
\end{figure}

In order to solve this problem, we introduce a threshold on the maximum
firing rates. We collect statistics of the peak positions of the firing
rate profile when their height exceeds the threshold. \textcolor{black}{The
result is shown in Figure \ref{fig:mean_sep}, indicating a significant
improvement of resolution compared with Figure \ref{fig:conditions}(D).}\textcolor{red}{{}
}\textcolor{black}{The effects of the threshold value on the resolution
performance are shown in Figure \ref{fig:Effect-of-threholding}.
When the threshold is low, the components are not resolved even at
a separation of 0.5 times the tuning width. On the other hand, when
the threshold is too high, the statistics of peak positions becomes
too sparse to be reliable. In an intermediate range of thresholds
that is not too narrow, the resolution of the components can be achieved
down to separations of 0.3 to 0.4 times the tuning width.}

\subsection{Recurrent Connections}

\begin{figure}
\begin{centering}
\includegraphics[width=0.8\textwidth]{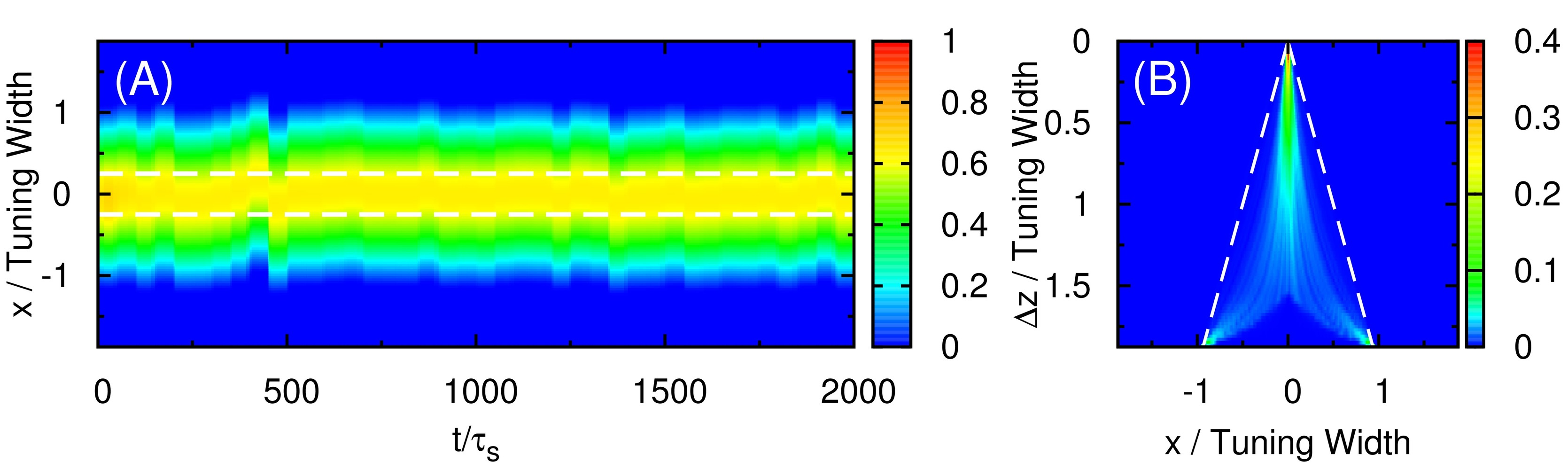}
\par\end{centering}

\caption{\label{fig:FeedForward}(A) Raster plot of firing rate $r$ for purely
feedforward network. $\Delta z=0.5$ TW. White dashed line: positions
of the two stimuli. (B) Contours of the distribution of peak positions
higher than $0.45$ as a function of preferred stimuli,$x$, and the
separation between the two stimuli, $\Delta z$. White dashed line:
positions of the two stimuli. Parameters: $J_{I}=0.3J_{E}$, $b=3a$,$\tau_{d}\beta/\rho J_{E}=0.2$,
$\rho J_{E}A=0.8$, $\sigma_{\delta A_{i}}/A_{0}=0.3$ and $\tau_{d}=50\tau_{s}$.}
\end{figure}

Finally, we would like to stress the importance of recurrent connections
in achieving resolution enhancement. With no recurrence, population
spikes cannot be generated and the amplification of the difference
between nearly overlapping inputs cannot be achieved. Let us consider
a purely feedforward network, with weaker but spatially broader inhibition
than excitation,

\begin{eqnarray}
\tau_{s}\frac{du}{dt}\left(x,t\right) & = & -u\left(x,t\right)+\rho\int dx'\left[J_{E}\exp\left(-\frac{\left|x-x'\right|^{2}}{2a^{2}}\right)-J_{I}\exp\left(-\frac{\left|x-x'\right|^{2}}{2b^{2}}\right)\right]p\left(x',t\right)I^{\text{ext}}\left(x',t\right)\nonumber \\
\label{eq:ff_u}\\
\tau_{d}\frac{dp}{dt}\left(x,t\right) & = & -p\left(x,t\right)+1-\tau_{d}\beta p\left(x,t\right)I^{\text{ext}}\left(x,t\right)\label{eq:ff_p}\\
r\left(x,t\right) & = & \Theta\left[u\left(x,t\right)\right]u\left(x,t\right),\label{eq:ff_r}
\end{eqnarray}
where $J_{E}>J_{I}$ and $I^{\text{ext}}\left(x,t\right)$ is the
same as that in recurrent network in Eq. \eqref{eq:Iext}. Although
in this feedforward network STD can still modulate the synaptic efficacy
so that neuronal activities prefer the side region to the midpoint
between two stimuli, temporal modulation, which is essential to population
spikes, cannot be realized without feedback. As mentioned above, population
spikes make it easier for the activity profile to switch off on one
side and grow up on the other. As shown in Figure \ref{fig:FeedForward},
the resolution enhancement in the purely feedforward network is poor.
In fact, the behavior is very similar to those in the non-spiking
region even when the architecture is recurrent, as shown in Figures
\ref{fig:phase_diagram}(A) and (E).

\section{{\normalsize Discussion}}

In this paper, we have demonstrated how STD plays the role of generating
population spikes that can carry information extra to spike rates.
We have used the example of resolving transparent motion with two
components in a continuous attractor neural network, and have shown
that the temporal modulation of the firing rates enables the network
to enhance the resolution of motion transparency, thereby providing
a possible explanation to the longstanding mystery of resolving separations
narrower than the tuning width of the neurons, and resulting in input-output
relations that can have excellent agreement with experimental results
\cite{Treue2000}. The role played by STD was further clarified by
comparison with alternate scenarios under 4 general conditions.

First, the strength of STD should be sufficiently strong. Weaker STD
may result in the network response being pinned by one of the two
components, or slosher modes that span a range of positions effectively
independent of the component separations. On the other hand, sufficiently
strong STD can give rise to population spikes, endowing them the freedom
to alternate between the two components. Equally important is the
provision of temporal modulation by the population spikes, so that
the firing patterns indeed contain information of the stimuli, even
though the time-averaged firing rate can only resolve separations
larger than the tuning width of neurons, as shown in Figure \ref{fig:average_activity}
and found experimentally by Treue \etal. \cite{Treue2000}. The role
played by temporally modulated signals in transparent motions can
be tested in future experiments.

Second, the strength of the input should be sufficiently strong. Otherwise,
no population spikes can be produced. Even for moderately strong input,
the population spikes become moving ones, and fail to represent the
stimulus positions.

Third, fluctuations in the input profiles are also important. They
provide the temporally sensitive signals when the two components cannot
be resolved in the time-averaged input. They correspond to the ``unbalanced
motion signals'' in the detection of transparent motion with opposite
moving directions \cite{Qian1994}.

Fourth, thresholds are needed to extract the information of the stimuli
contained in the firing patterns, since they are able to truncate
background activities that interfere the signals from the two components.

Our proposed model is not the first model or mechanism to explain
the behavior of the discriminational task in transparent motion experiments.
It was suggested that the curvature of the average neural activity
may provide information of multiple stimuli, but the neural activity
is wider than expected \cite{Treue2000}. Other proposals require
more complex structures to achieve the task. For example, a population
to encode uncertainty is needed to differentiate between multiplicity
and uncertainty \cite{Sahani2003}, and additional internal structures
are needed to provide feedback information \cite{Raudies2011}. While
admittedly involving additional structures and layers can augment
the functionality of the brain, our work shows that it is possible
to achieve with little additional structure the performance consistent
with experiments in \cite{Treue2000} and \cite{Braddick2002}. An
interesting future direction is to consider whether firing rates multiplexed
with temporal modulations can be an instrument to achieve the differentiation
between multiplicity and uncertainty posed in \cite{Sahani2003}.

The ability of STD to generate temporally modulated response is also
applicable to other brain tasks, such as switching between percepts
in competitive neural networks \cite{Kilpatrick2012}. Compared with
other conventional neural network models processing time-averaged
or static neuronal response profiles, the temporal component provides
an extra dimension to encode acute stimuli, so that information processing
performance can be significantly enhanced.

\subsubsection*{Acknowledgment}

This work is supported by the Research Grants Council of Hong Kong
(grant numbers 605010, 604512 and N\_HKUST606/12) and the National
Foundation of Natural Science of China (No. 31221003, No.31261160495). 

\newpage{}

\section*{Appendix}

\begin{figure}[H]
\begin{centering}
\includegraphics[width=0.65\textwidth]{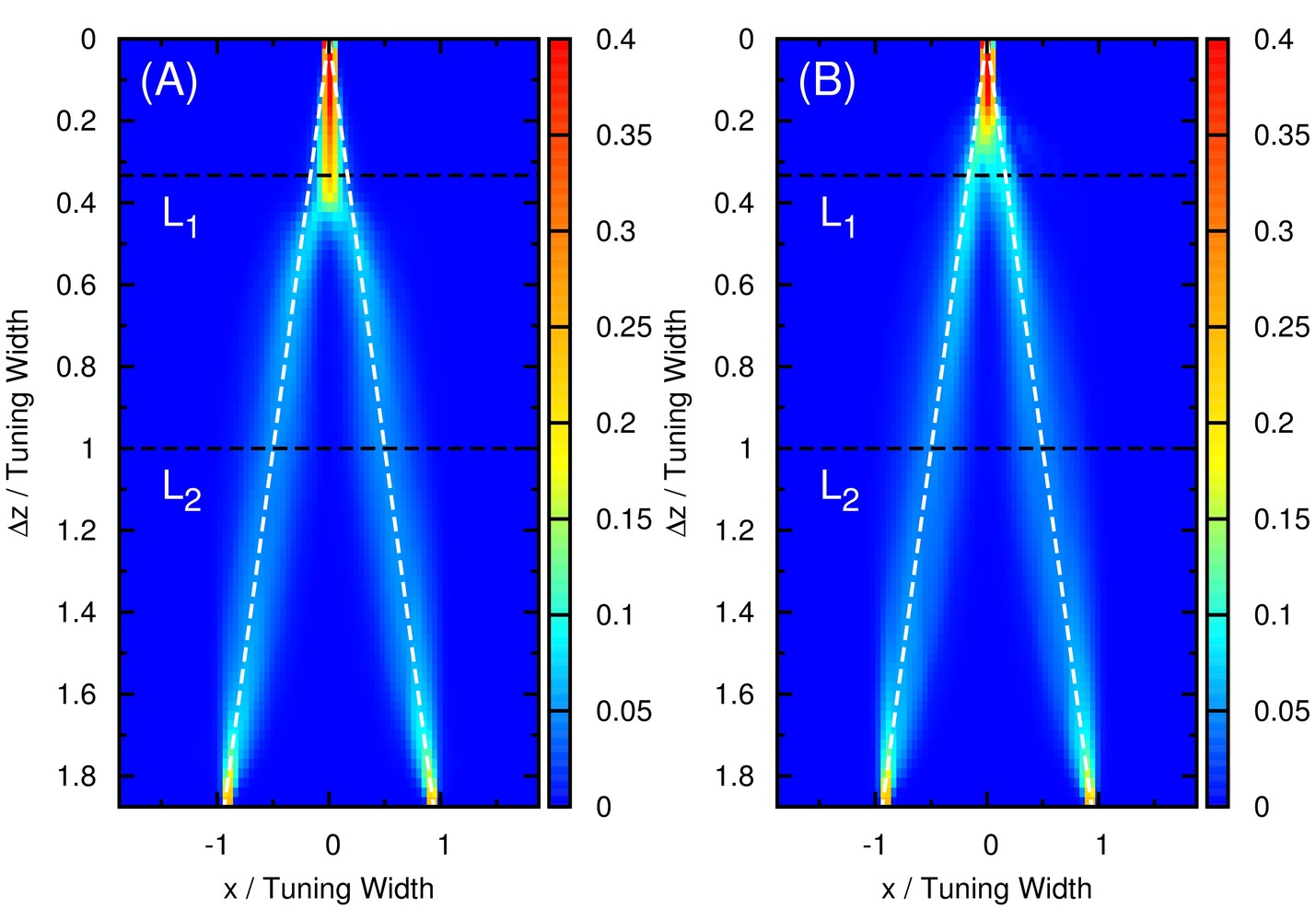} 
\par\end{centering}

\begin{centering}
\caption{\label{fig:prob_den_cutoff_var_aI}Contours of the distribution of
peak positions higher than (A) 6.2 and (B) 5.5 as a function of preferred
stimuli, $x$, and the separation between the two stimuli, $\Delta z$.
White dashed line: positions of two stimuli. Parameters: same as Figure
\ref{fig:network_act_var_dz}, except $a_{I}=0.95a$ for (A) and $a_{I}=1.05a$
for (B).}
\ \\
\includegraphics[width=0.65\textwidth]{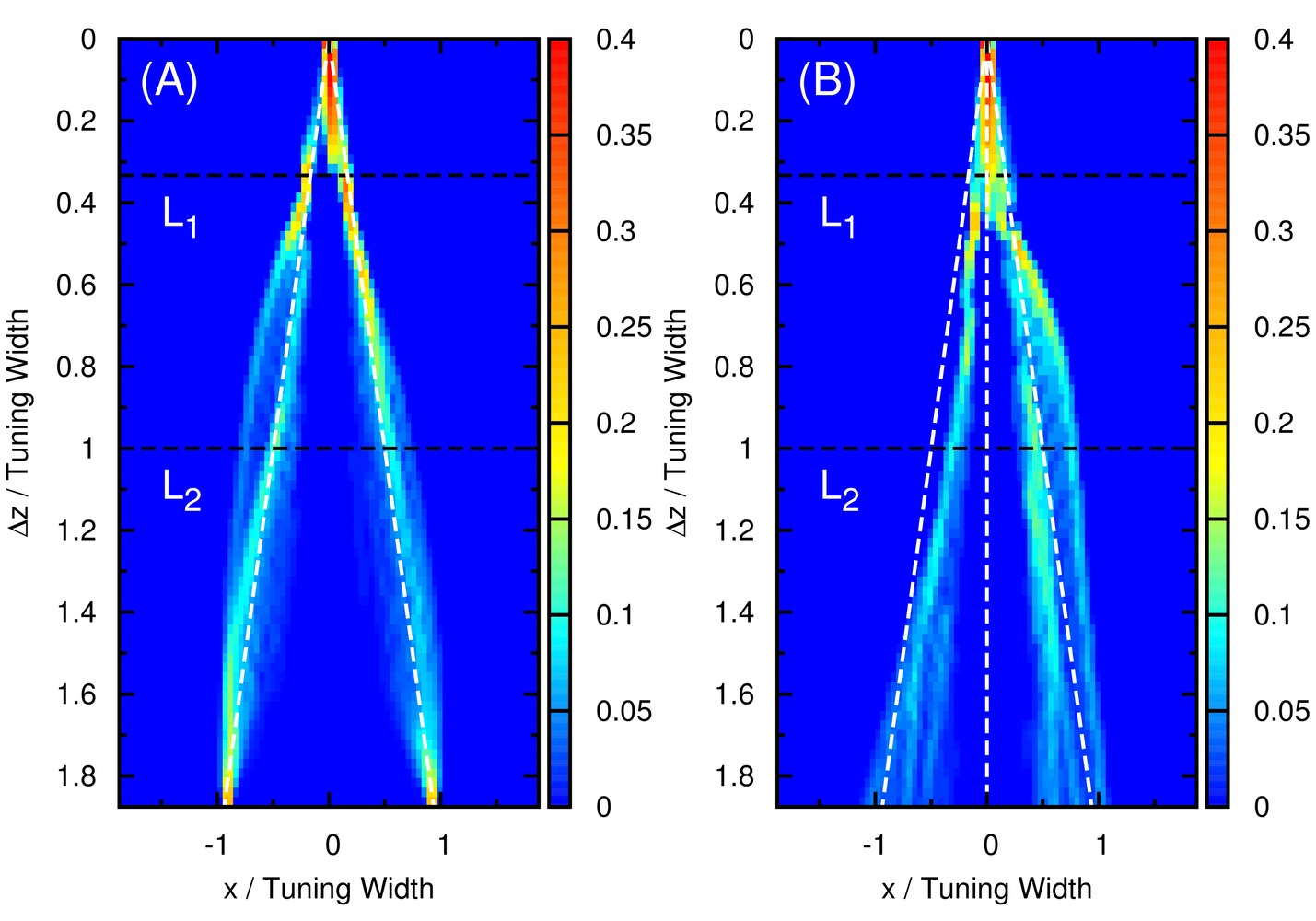} 
\par\end{centering}

\caption{\label{fig:fast_decode}The population spike occurrence counted within
500$\tau_{s}$, which is comparable to the timescale in typical experiments.
(A) Situations that there are two stimuli. (B) Situations with three
stimuli. Other parameters: (A) same as Figure \ref{fig:average_activity},
(B) same as Figure \ref{fig:prob_den_cutoff_3sti}. }
\end{figure}

\end{document}